\documentclass[reprint,nofootinbib]{revtex4-2}
\usepackage{stmaryrd}
\usepackage{preamble}

\begin{document}
\newcommand{\Rn}{\mathbb{R}^n}

\title{The Ricci decomposition of the inertia tensor for a rigid body in arbitrary spatial dimensions}
\author{Edward Parker}
\email{tparker@alumni.physics.ucsb.edu}
\date{\today}

\begin{abstract}
The rotations of rigid bodies in Euclidean space are characterized by their instantaneous angular velocity and angular momentum. In an arbitrary number of spatial dimensions, these quantities are represented by bivectors (antisymmetric rank-2 tensors), and they are related by a rank-4 inertia tensor. Remarkably, this inertia tensor belongs to a well-studied class of \emph{algebraic curvature tensors} that have the same index symmetries as the Riemann curvature tensor used in general relativity. Any algebraic curvature tensor can be decomposed into irreducible representations of the orthogonal group via the \emph{Ricci decomposition}. We calculate the Ricci decomposition of the inertia tensor for a rigid body in any number of dimensions, and we find that (unlike for the Riemann curvature tensor) its Weyl tensor is always zero, so the inertia tensor is completely characterized by its (rank-2) Ricci contraction. So unlike in general relativity, the Weyl tensor does not cause any qualitatively new phenomenology for rigid-body dynamics in $n \geq 4$ dimensions.
\end{abstract}

\maketitle

\section{Introduction}

A first course in classical mechanics usually begins by discussing the rotation of rigid plane figures in two dimensions. In this context, students are taught the familiar equations for the 2D rigid-body rotation about a fixed point
\beq \label{2D}
L = I^{(2D)} \omega,\quad \tau = \frac{dL}{dt} = I^{(2D)} \alpha,\quad T = \frac{1}{2} I^{(2D)} \omega^2,
\eeq
where $L$ is the body's angular momentum, $I^{(2D)} = \int dm\, r^2$ is its moment of inertia about the point of rotation ($dm = \sigma(\bm{r})\, d^2r$ with $\sigma(\bm{r})$ the area mass density), $\omega$ is its angular velocity, $\tau$ is the net external torque $\sum r\, F_\theta$ applied to it, $t$ is time, $\alpha$ is the body's angular acceleration, and $T$ is its rotational kinetic energy. In the 2D context, $I^{(2D)}$ and $T$ are considered to be nonnegative scalars and $L$, $\omega$, $\tau$, and $\alpha$ to be signed scalars whose signs represent a counterclockwise or clockwise orientation. Students are sometimes vaguely told that similar formulas often work for 3D rotation about axes with high symmetry and given somewhat mysterious formulas for the moments of inertia about various axes for various 3D shapes.

A later course will usually cover the rotation of rigid bodies in 3D more systematically. Students learn that in 3D, equations~\eqref{2D} generalize to
\beq \label{3D}
\bm{L} = I^{(3D)} \bm{\omega}, \ \bm{\tau} = \frac{d\bm{L}}{dt}, \ T = \frac{1}{2} \bm{\omega} \cdot I^{(3D)} \cdot \bm{\omega} := \frac{1}{2} \omega_i I^{(3D)}_{ij} \omega_j
\eeq
(in the inertial ``laboratory'' frame of reference) \cite{Thornton}. The kinetic energy $T$ remains a scalar quantity, but $\bm{L}$, $\bm{\omega}$, $\bm{\tau}$, and $\bm{\alpha}$ are now considered to be \emph{vector} quantities. Now $\bm{L} = \int \bm{r} \cross dm\, \bm{v}(\bm{r})$ (where now $dm = \rho(\bm{r}) d^3r$ with $\rho(\bm{r})$ the volume mass density), $\bm{\omega}$ is oriented along the axis of rotation with a magnitude equal to the angular speed, $\bm{\tau} = \int \bm{r} \cross d\bm{F}_\text{ext}(\bm{r})$, and $\bm{\alpha} = d\bm{\omega}/dt$. (More precisely, they are all pseudovector quantities that do not change orientation under a parity inversion.) The scalar moment of inertia $I^{(2D)}$ generalizes to a symmetric rank-2 inertia \emph{tensor}, or (more prosaically but concretely) a symmetric $3 \times 3$ matrix $I^{(3D)}$. The components of this tensor are determined by the rigid body's mass distribution:
\beq \label{I3D}
I^{(3D)}_{ij} = \int dm \left( r^2 \delta_{ij} - r_i r_j \right).
\eeq
Since the tensor is represented by a real symmetric matrix, it can always be diagonalized. Its eigenvectors are referred to as the rigid body's \emph{principal axes} and the corresponding eigenvalues are the \emph{principal moments of inertia} about those axes. Only for rotations about the principal axes do the vector equations~\eqref{3D} simplify to the scalar equations~\eqref{2D}.

The tensor $I^{(3D)}$ is no longer necessarily constant in the laboratory frame as its orientation changes, so the easiest course of action is often to shift to a non-inertial ``body'' frame of reference in which $I^{(3D)}$ is constant, even at the expense of the additional complications from working in a non-inertial reference frame. Any student who has studied rigid-body motion in 3D can testify that the relatively simple equations~\eqref{3D} and \eqref{I3D} can already lead to very complicated and unintuitive rotational dynamics.

But three dimensions are not the end of the story. What about an arbitrary number of dimensions $n$? Rigid-body rotation in higher than three dimensions is admittedly less realistic than $n=2$ or $n=3$, but considering the fully general case is still an interesting thought exercise that will yield unexpected connections to the study of general relativity, which \emph{is} naturally formulated in greater than three dimensions.

Although this article will eventually use some advanced tools developed to study general relativity, most of it should be accessible to someone with a solid understanding of advanced undergraduate classical mechanics. We leave some of the more technical mathematical details in the footnotes.

\section{Mathematical preliminaries}

(This section can be skipped by those who are less concerned with the mathematical details.)

We define a \emph{rigid body} to be a set of points whose relative distances remain constant. An extended body can only remain rigid if the causal influence of an external force on one point is instantly transmitted to all other points, so rigid bodies cannot exist in a relativistic setting. Moreover, we will require that the body is free to perform unconstrained rotation, which requires that space be flat. We will therefore work in the Euclidean space $\Rn$ endowed with the usual flat and positive-definite Euclidean inner product (except where indicated otherwise).\footnote{In this article, the notation $\Rn$ always refers to the full Euclidean inner product space, not just the vector space.} We will always work in Cartesian coordinates, in which the metric indices are given by the Kronecker delta $\delta_{ij}$. We will use the Einstein summation convention that repeated indices are summed from 1 to $n$, and we will not distinguish between raised and lowered tensor indices. We will denote the (constant) Euclidean metric tensor by $\delta$ and will only occasionally use $g$ to denote the metric tensor \emph{field} for an arbitrary (potentially curved) manifold.

\subsection{The exterior algebra}

Several ideas from the \emph{exterior algebra} will be very useful \cite{Nakahara}. If $k$ is a natural number, then a \emph{$k$-vector} or \emph{multivector} is an element of $\Lambda^k(V)$, the $k$th exterior power of a vector field $V$.\footnote{We use the terminology convention that ``multivectors'' must be homogeneous with fixed $k$. The terms ``multivector'', ``bivector'', etc.\ are often associated with the somewhat obscure formalism of \emph{geometric algebra}, but this article does not use any concepts from geometric algebra -- just the simpler and much more standard exterior algebra of totally antisymmetric tensors.  Some physicists familiar with general relativity might be more used to referring to totally antisymmetric tensors as ``differential forms''. But strictly speaking, differential forms are smooth multivector \emph{fields} that are functions of a spacetime manifold. The bivectors discussed in this article are not local fields but correspond to individual extended objects, so they are just fixed bivectors and \emph{not} differential forms.} At a ``physicist's level of rigor'', $\Lambda^k(V)$ is the space of totally antisymmetric rank-$k$ tensors over $V$. A $p$-vector $A$ and a $q$-vector $B$ can be combined together into a $(p+q)$-vector using the \emph{wedge product}
\[
(A \wedge B)_{\mu_1, \dots, \mu_{p+q}} = \frac{(p+q)!}{p!\, q!} A_{[\mu_1 \dots \mu_p}B_{\mu_{p+1} \dots \mu_{p+q}]},
\]
where $[\,]$ around tensor indices denotes total antisymmetrization.\footnote{There are two different normalization conventions for the wedge product in common use. In this article, we use the ``geometer's convention'' that is standard in physics rather than the ``algebraist's convention''. See \cite{MO} for a detailed discussion of the pros and cons of each convention.} A multivector is \emph{simple} (or \emph{decomposable} or a \emph{$k$-blade}) if it can be expressed as a wedge product $v_1 \wedge \dots \wedge v_k$ of $k$ rank-1 vectors $v_i$.

$\Lambda^k(\Rn)$ is a real vector space of dimension $\binom{n}{k}$, since a natural basis is the set of simple wedge products $e_{i_1} \wedge \dots \wedge e_{i_k}$ of $k$ unit vectors within an orthonormal basis for $\Rn$. Moreover, we can use the inner product on $\Rn$ to map any multivector $A \in \Lambda^k(V)$ to its \emph{Hodge dual} multivector $\star A \in \Lambda^{n-k}$. For Euclidean space, the Hodge dual of a multivector is just proportional to its contraction with the totally antisymmetric Levi-Civita tensor $\epsilon_{\mu_1 \dots \mu_n}$: 
\[
(\star A)_{\mu_1 \dots \mu_{n-k}} = \frac{1}{k!} \epsilon_{\mu_1 \dots \mu_{n-k} \nu_1 \dots \nu_{k}} A_{\nu_1 \dots \nu_k}.
\] 
(The formula is more complicated for more general manifolds.) For a Reimannian (i.e. positive-definite) metric, the double Hodge star of a $k$-vector $A \in \Lambda^k(\Rn)$ is $\star \star A = (-1)^{k(n-k)} A$. The vector space $\Lambda^k(\Rn)$ inherits its own inner product from the Euclidean inner product. The inner product between simple $k$-vectors is given by
\[
\langle v_1 \wedge \dots \wedge v_k, w_1 \wedge \dots \wedge w_k \rangle := \det M, \ \ M_{ij} := \langle v_i, w_j \rangle,
\]
and it extends to general multivectors by linearity. As a special case, the norm-squared of a simple $k$-vector $v_1 \wedge \dots \wedge v_k$ equals the \emph{Gram determinant} of the matrix with entries $\langle v_i, v_j \rangle$.\footnote{We will use the normalization convention that when calculating the inner product of $k$-vectors, the combinatorial factor $1/k!$ goes into the index contraction rather than into the antisymmetric tensors that represent the orthonormal basis vectors $\hat{e}_{i_1} \wedge \dots \wedge \hat{e}_{i_k}$ for $\Lambda^k(\Rn)$. That is, the tensors that represent $\hat{e}_{i_1} \wedge \dots \wedge \hat{e}_{i_k}$ have elements $1$, $0$, and $-1$ for all $k$, while the inner product on the exterior algebra $\Lambda^k(\Rn)$ is given by $\langle A, B\rangle_{\Lambda^k(\Rn)} = \frac{1}{k!} A_I B_I$ (where $I$ denotes the multi-index $(i_1, \dots, i_k)$) instead of by the usual inner product $A_I B_I$ on the tensor algebra.}

\subsection{Rotations in arbitrary dimensions}

By definition, a rotation $R$ of $n$-dimensional Euclidean space preserves angles and distances between points, and more generally it preserves the Euclidean inner product between vectors. It is also straightforward to show that a rotation must be a linear transformation on vectors. Therefore, for any vectors $v$ and $u$, $\delta(v, u) \equiv \delta(Rv, Ru)$. In matrix language, this becomes
\[
(Rv)^T \delta (Ru) = v^T R^T \delta R u = v^T \delta u,
\]
where $\delta$ represents the $n \times n$ identity matrix. Since this equation must hold for all vectors $u$ and $v$, we must have that $R^T \delta R = \delta$. If we only consider proper rotations, which are connected to the identity operator, then the set of proper rotation operators form the Lie group $\mathrm{SO}(n)$.

Angular velocity and angular momentum correspond to infinitesimal rotations, which are elements of the Lie algebra $\mathfrak{so}(n)$. To see what the elements $A$ of (the fundamental representation of) $\mathfrak{so}(n)$ look like, we can Taylor expand the rotation operator in equation $R^T R = \delta$ to first order in the rotation angle $\theta$: letting $R = \delta + \theta A + o(\theta^2)$ gives
\begin{align*}
\left( \delta + \theta A^T + o(\theta^2) \right) & \left( \delta + \theta A + o(\theta^2) \right) \\ &= \delta + \theta \left(A + A^T \right)+ o(\theta^2) \\
&= \delta,
\end{align*}
so $A + A^T = 0$ and A must be an $n \times n$ antisymmetric matrix. An infinitesimal rotation generator $A \in \mathfrak{so}(n)$ can be mapped to a rotation $R \in \mathrm{SO}(n)$ through a non-infinitesimal angle $\theta$ by the \emph{exponential map}  $R = \exp(\theta A)$.\footnote{In the context of quantum mechanics, physicists usually use the convention that generators are Hermitian operators and the exponential map is given by $A \to \exp(-i A t/\hbar)$, where $t$ is a continuous real parameter like time, distance, or angle. In this article, it will be easier to stick to real numbers and use the phase convention more common among mathematicians.} For matrix representations like the one that we are implicitly considering, the exponential map is just the ordinary matrix exponential.

At our level of rigor, antisymmetric matrices are \emph{bivectors} in $\Lambda^2(\Rn)$. So in general dimensions, an infinitesimal rotation is not represented by a (pseudo-)vector but by a bivector \cite{Jensen}.\footnote{The Lie algebra $\mathfrak{so}(n)$ is isomorphic to $\Lambda^2(\mathbb{R}^n)$ as a vector space, but instead of the wedge product it has a \emph{Lie bracket} given by the matrix commutator. We will not need this Lie bracket in this article.} 

More concretely, if $x_i$ and $x_j$ are orthonormal vectors in $\Rn$, then an infinitesimal rotation in the $i$-$j$ plane (oriented so that $x_i$ rotates into $x_j$) is generated by the simple bivector $x_i \wedge x_j$. More generally, the magnitude of a simple bivector gives the (infinitesimal) angle of rotation, and its sign (or equivalently, the ordering of the two vectors being wedged together) gives the orientation of the rotation.\footnote{When we say that rotations are ``infinitesimal'', we mean that they are small enough that we can neglect any small non-commutative composition effects and add them together without keeping track of ordering. The higher-order non-commutative effects are captured by the Lie bracket structure mentioned in a previous footnote.} A \emph{simple} bivector generates a rotation in a single plane that leaves all orthogonal directions unchanged. (In $n>3$ dimensions, we cannot describe this as a rotation about a single 1D axis, because there are multiple directions that are all orthogonal to the plane and to each other.) All bivectors over $\Rn$ are simple if $n \leq 3$, so all rotations occur in a single plane. But not all bivectors are simple if $n \geq 4$; instead, any bivector in $\Lambda^2(\Rn)$ can be decomposed into a sum of at most $\left \lfloor \frac{n}{2} \right \rfloor$ orthogonal simple bivectors (where $\lfloor\, \rfloor$ denotes the floor function). This decomposition is generically unique, unless multiple simple bivectors have the same magnitude \cite{Jensen}.\footnote{In matrix language, a real antisymmetric $n \times n$ matrix $A$ represents a \emph{simple} bivector iff there exists a nonzero vector $v \in \Rn$ such that $A_{[ij} v_{k]} \equiv 0$. This is true for all real antisymmetric $n \times n$ matrices $A$ if $n \leq 3$, but only for some such matrices if $n \geq 4$. Geometrically, it means that the vector $v$ lies in the unique plane in $\Rn$ spanned by the simple bivector $A$.} Therefore, not all rotations of higher-dimensional Euclidean space occur in a single plane; a general rotation of Euclidean space in $n$ dimensions is generated by orthogonal planes rotating simultaneously (generically at different speeds).

In three dimensions, the usual (pseudo-)vector representations of angular velocity, angular acceleration, angular momentum, and torque are all derived from the fundamental infinitesimal pseudovector rotation $d\bm{\theta}$. These pseudovector representations are the Hodge duals of the corresponding bivectors. (The fact that they transform as pseudovectors is a clue that a bivector description is more fundamental, because unlike pseudovectors, bivectors transform in the natural way under parity inversion \cite{Jensen}.) The bivector representations of all of these quantities are defined in any dimension, but the pseudovector representations only make sense for $n=3$. As expected, for $n=2$ the bivector space is one-dimensional, representing the single scalar degree of freedom for plane rotations. For $n = 3$, the bivector space is three-dimensional, corresponding to the usual axis-magnitude representation of a 3D rotation. But for $n = 4$, the bivector space is \emph{six}-dimensional -- more than the four degrees of freedom that we might expect based on our 3D intuition.

\section{The inertia tensor in arbitrary dimensions}

In three dimensions, the inertia tensor is a linear map that maps an angular velocity \mbox{(pseudo-)}vector to an angular momentum (pseudo-)vector. But in arbitary dimensions, angular velocity and angular momentum are represented by bivectors, not vectors. The inertia tensor therefore generalizes to a rank-4 tensor $I:\Lambda^2(\Rn) \to \Lambda^2(\Rn)$ that linearly maps bivectors to bivectors. In terms of indices, this becomes\footnote{The factor of $1/2$ in \eqref{L} is not necessary. We simply include it to match the standard normalization for the 3D rank-2 tensor, and to parallel our conventions for the inner product on $\Lambda^k(\Rn)$ and the Hodge star operator that we normalize a contraction of $k$ totally antisymmetric tensor indices by $1/k!$. But the tensor contraction in \eqref{L} does not represent an inner product on $\Lambda^k(\Rn)$, so we could also consistently absorb the factor of $1/2$ into the normalization of $I$.}
\beq \label{L}
L_{ij} = \frac{1}{2} I_{ijkl} \omega_{kl}.
\eeq
More abstractly, the inertia tensor can still be thought of as a linear operator on a real inner product space -- but the inner product space is no longer the $n$-dimensional physical Euclidean space, but the $\binom{n}{2} = \frac{1}{2}n(n-1)$-dimensional inner product space $\Lambda^2(\Rn)$ of bivectors on $\Rn$. This linear operator will turn out to be self-adjoint, just like in the 3D case. 

After all this setup, it is actually very simple to derive the inertia tensor. In arbitrary dimensions, we do not have a cross product, so the angular momentum generalizes to a bivector wedge product of the vectors $\bm{r}$ and $dm\, \bm{v}$:
\[
L = \int \bm{r} \wedge (dm\, \bm{v}).
\]
($dm$ now represents the arbitrary-dimensional volume form $\rho(\bm{r})\, d^nr$.) With our choice of sign conventions, the 3D rigid-rotation formula $\bm{v} = \bm{\omega} \cross \bm{r}$ (which ultimately derives from $d\bm{r} = d\bm{\theta} \cross \bm{r}$) generalizes to $v_j = r_k \omega_{kj}$. So
\begin{align}
L_{ij} &= \int dm \left( 2 r_{[i} v_{j]} \right) 
= \int dm \left( 2 r_{[i|} r_k \omega_{k|j]} \right) \label{Ldirect} \\
&= \int dm \left( 2 r_{[i|} r_k \delta_{|j]l} \omega_{kl} \right) \nn \end{align}

This equation would seem to suggest that $I_{ijkl} = \int dm \left( 4 r_{[i} \delta_{j]l} r_k \right)$. Strictly speaking, this formula is correct in the sense that it returns the correct value of $L$, but it contains unphysical degrees of freedom. The inertia tensor inputs a bivector angular momentum $\omega$ that is always antisymmetric, so any part of $I_{ijkl}$ that is symmetric in $k$ and $l$ will vanish by symmetry when contracted with $\omega_{kl}$, and only the part that is antisymmetric in $k$ and $l$ will affect the output $L$. We therefore explicitly antisymmetrize $I$ on $k$ and $l$ to more clearly show which are the true degrees of freedom that affect the angular momentum:
\begin{align}
\Aboxed{I_{ijkl} &= \int dm\, \left( -4 r_{[i} \delta_{j][k} r_{l]} \right)} \label{I} \\
&= \int dm\, \left( -r_i \delta_{jk} r_l + r_i \delta_{jl} r_k + r_j \delta_{ik} r_l - r_j \delta_{il} r_k \right). \nonumber
\end{align}

In three dimensions, \eqref{L} can be reformulated in terms of pseudovector quantities as
\[
\bm{L} = \star L = \star \left( \frac{1}{2} I \omega \right) = \star \left( \frac{1}{2} I(\star \bm{\omega}) \right),
\]
or in index notation,
\[
L_p = \frac{1}{4} \epsilon_{pij} I_{ijkl} \epsilon_{klq} \omega_q.
\]
Comparing with the first equation in \eqref{3D}, we see that
\[
I^{(3D)}_{pq} = \frac{1}{4} \epsilon_{pij} I_{ijkl} \epsilon_{klq} = \int dm \left( r^2 \delta_{pq} - r_p r_q \right),
\]
which agrees with \eqref{I3D}.

The inertia tensor satisfies several index symmetries:
\begin{subequations} \label{ids}
\begin{align}
I_{ijkl} &= -I_{jikl} = -I_{ijlk} \label{anti} \\
I_{ijkl} &= I_{klij} \label{sym} \\
I_{ijkl} &+ I_{iklj} + I_{iljk} = 0. \label{Bianchi}
\end{align}
\end{subequations}
Remarkably, these are the exact same symmetries satisfied by the Riemann curvature tensor in general relativity \cite{Berrondo}. But there is one important structural difference between the inertia tensor and the Riemann curvature tensor: the inertia tensor is a single fixed tensor, while the Riemann curvature tensor is a tensor \emph{field} defined over a spacetime manifold.\footnote{There is a subtle point here. The integrated tensor $I$ is indeed just a single tensor with no spatial dependence. But, as mentioned above, the differential $dm$ is technically a true differential volume \emph{form} on $\Rn$ (although the full machinery of differential forms is somewhat overkill for integrating over Euclidean space). The integrand in parentheses in \eqref{I} explicitly depends on the position $\bm{r}$ and so is obviously a tensor \emph{field} that varies over space. Taken together, the full differential form being integrated in \eqref{I} is a \emph{tensor-valued volume form} \cite{Bini}. Tensor-valued differential forms cannot be integrated over generic curved spaces, because there is no natural way to parallel-transport the tensor at each point in the manifold to the same base point so that the tensors can be added together within the same vector space. But vector-valued differential forms defined on a flat manifold \emph{can} be integrated, because the vector spaces at each point are naturally isomorphic.} The last section of this article exploits this parallel by using tools from general relativity to study the inertia tensor for rigid bodies in Euclidean spacetime.

\section{The Ricci decomposition of the inertia tensor}

\subsection{Algebraic curvature tensors}

Any rank-4 tensor that satisfies the index symmetries \eqref{ids}, including the inertia tensor $I$ given by \eqref{I}, is referred to as an \emph{algebraic curvature tensor} by analogy with the Riemann curvature tensor \cite{Besse}.

Identity \eqref{Bianchi} is known as the \emph{first} or \emph{algebraic Bianchi identity}.\footnote{In keeping with Stigler's law of eponymy, the algebraic Bianchi identity was discovered by Ricci.} It follows from identities~\eqref{anti} and  \eqref{sym} if $n = 2$ or $3$, but is an independent condition if $n \geq 4$ \cite{Besse}. Identity~\eqref{anti} simply means that an algebraic curvature tensor can be thought of as a linear operator on $\Lambda^2(\Rn)$. Identity~\eqref{sym} means that this operator is self-adjoint, just as the rank-2 inertia tensor is in 3D. Therefore, there always exists a complete orthonormal basis of $\binom{n}{2}$ eigenbivectors $\omega^{(i)}$ with eigenvalues $I^{(i)}$ (the principal moments of inertia) such that if the rigid body is rotating with angular velocity $\omega^{(i)}$, then its angular momentum $L = I^{(i)} \omega^{(i)}$. These eigenbivectors are the generalizations of the principal axes in 3D (but in higher dimensions, they may not be simple and so may not correspond to rotations within a single plane). In general relativity, this approach (thinking of the Riemann curvature tensor as a self-adjoint operator on $\Lambda^2(M)$ and considering its eigendecomposition) leads to the \emph{Petrov classification} of spacetimes \cite{Petrov}.

\subsection{The Kulkarni-Nomizu product}

It will be convenient to introduce the bilinear \emph{Kulkarni-Nomizu product} of symmetric rank-2 tensors \cite{Besse}. If $A$ and $B$ are symmetric rank-2 tensors, then their Kulkarni-Nomizu product $A \owedge B$ is a rank-4 tensor defined by
\[
(A \owedge B)_{ijkl} := A_{ik} B_{jl} - A_{il} B_{jk} - A_{jk} B_{il} + A_{jl} B_{ik}.
\]
Any Kulkarni-Nomizu product is an algebraic curvature tensor that satisfies \eqref{ids} (although the converse is not true).\footnote{There is a \emph{formal} similarity between the commutation relations $[J^{\mu \nu}, J^{\rho \sigma}]$ for the Lie algebra $\mathfrak{so}(3,1)$ (the generators of the $(3+1)D$ Lorentz group) and the formal Kulkarni-Nomizu product $-i\, \eta \owedge J$, where $\eta$ is the flat Minkowski metric with signature $(-,+,+,+)$ (and we use the standard physicists' convention for the normalization of the generators, rather than the mathematicians' convention used in the main text) \cite{Srednicki}. But the latter expression is not actually a true Kulkarni-Nomizu product, because the tensor $J$ is antisymmetric rather than symmetric. The commutator is therefore not an algebraic curvature tensor; it is antisymmetric rather than symmetric under the simultaneous exchange $(\mu \leftrightarrow \rho,\ \nu \leftrightarrow \sigma)$, and requirement~\eqref{sym} is violated.} The Kulkarni-Nomizu product is symmetric: $A \owedge B \equiv B \owedge A$.

Note that
\[
(A \owedge A)_{ijkl} = 4 A_{i[k} A_{l]j}.
\]
For an arbitrary two-dimensional curved surface, the Riemann curvature tensor field equals $\frac{1}{4} R\, g \owedge g$, where $R$ is the Ricci scalar field (twice the Gaussian curvature) and $g$ is the metric tensor field. All \emph{space forms} -- Riemannian manifolds of any dimension with constant \emph{sectional curvature} -- also have a Riemann curvature tensor field equal to $\frac{1}{4} R\, g \owedge g$, although in this case the Ricci scalar field $R$ is constant over the manifold.

Also note that if $v$ is a rank-1 vector, then
\[
(A \owedge (v \otimes v))_{ijkl} = -4 v_{[i} A_{j][k} v_{l]}.
\]
Equation~\eqref{I} therefore simplifies to the compact expression
\beq
\boxed{I = \delta \owedge \int dm\, (\bm{r} \otimes \bm{r}).} \label{IKN}
\eeq

\subsection{The Ricci decomposition}

If $R$ is either an algebraic curvature tensor on $\Rn$ or an algebraic curvature tensor \emph{field} on an arbitrary $n$-dimensional manifold with metric tensor $g$, then $R$ has only one independent single trace: the rank-2 symmetric \emph{Ricci-contracted} tensor (or tensor field)\cite{Besse}\footnote{If we consider the linear map from the space of symmetric rank-2 tensors $A$ to the space of algebraic curvature tensors that is given by $A \to g \owedge A$ (where $g$ is an arbitrary inner product on $\Rn$), then this map turns out to be exactly the transpose of the Ricci contraction.}
\[
R^{(2)}_{jl} := R_{ijil}.
\]
The other five single traces equal $\pm R_{jl}$ or 0. The only independent double trace of $R$ is the Ricci-contracted scalar (or scalar field)
\[
R^{(0)} := R^{(2)}_{ii} = R_{ijij}.
\]
From these, we can form the trace-free Ricci-contracted tensor (or tensor field)
\[
\hat{R}^{(2)} := R^{(2)} - \frac{1}{n} R^{(0)} g.
\]

$R$ can be naturally decomposed into a (direct) sum of three irreducible representations of the orthogonal group $\mathrm{O}(n)$:
\beq \label{decomp}
R = S + E + C.
\eeq
This decomposition is known as the \emph{Ricci decomposition} \cite{Besse}. Here $S$, $E$, and $C$ are themselves algebraic curvature tensors (or tensor fields) given by
\begin{align*}
S &:= \frac{R^{(0)}}{2n(n-1)} g \owedge g \\
E &:= \frac{1}{n-2} \hat{R}^{(2)} \owedge g \\
C &:= R - S - E.
\end{align*}
For $n = 2$, only the $S$ term is well defined, so the decomposition is trivial. For $n = 3$, the $C$ term vanishes identically. For $n \geq 4$, all three terms are generically nonzero.

Equation \eqref{decomp} is mathematically trivial by the definition of $C$, but the Ricci decomposition is useful because each term lies in a different irreducible representation of the orthogonal group. Loosely speaking, the $S$ term contains the doubly-contracted degree of freedom in the algebraic curvature tensor that transforms under rotations as a scalar, the $E$ term contains the singly-contracted degrees of freedom that transform under rotations as a traceless symmetric rank-2 tensor, and the $C$ term contains the uncontracted degrees of freedom that transform under rotations as a rank-4 tensor.

In general relativity, the $S + E$ terms contain the same information as the Ricci tensor field, which reflects spacetime's local response to matter, while $C$ is the Weyl tensor field, which is totally traceless and contains the gravitational degrees of freedom that propagate through vacuum. We can simplify the former sum to
\beq
S + E = A \owedge g, \label{SE}
\eeq
where the symmetric rank-2 tensor
\begin{align*}
A :=&\ \frac{1}{n-2} \hat{R}^{(2)} + \frac{1}{2n(n-1)} R^{(0)} g \\
=&\ \frac{1}{n-2}\left( R^{(2)} - \frac{1}{2(n-1)} R^{(0)} g \right)
\end{align*}
is known as the \emph{Schouten tensor} for the algebraic curvature tensor \cite{Kuhnel}.\footnote{Some sources normalize the Schouten tensor to be twice this expression or to have the opposite sign.} The Schouten tensor and the Ricci tensor are very closely related, and either can be easily derived from the other; they are essentially just rescaled and trace-adjusted versions of each other.

For the fixed inertia tensor \eqref{I}, we have (shifting notation from $R$ to $I$)
\begin{align}
I^{(2)} &= \int dm \left( (n-2) \bm{r} \otimes \bm{r} + r^2 \delta \right) \nonumber \\
I^{(0)} &= 2(n-1) \int dm \left( r^2 \right) \nn \\
\hat{I}^{(2)} &= (n-2) \int dm \left( \bm{r} \otimes \bm{r} - \frac{1}{n} r^2 \delta \right) \nn \\
A &= \int dm\, (\bm{r} \otimes \bm{r}). \label{A}
\end{align}
Combining equations~\eqref{SE}, \eqref{A}, and \eqref{IKN}, we see that $I = S + E$ and so $C = 0$. Unlike for the Riemann curvature tensor in general relativity, the Weyl component of the inertia tensor for rigid-body rotation vanishes identically in \emph{all} dimensions.

The inertia tensor $I$ is fully characterized by its Schouten tensor~\eqref{A} via equation~\eqref{IKN}. For $n \geq 3$ dimensions, the inertia tensor only has the $\frac{1}{2} n (n+1)$ independent degrees of freedom of its Schouten tensor. For $n \geq 4$, this is less than the $\frac{1}{12}n^2 (n^2 - 1)$ degrees of freedom in a generic algebraic curvature tensor.

Moreover, the fact that the Weyl part of $I$ vanishes follows directly from equation~\eqref{IKN} by rotational symmetry. $I$ depends only on a rank-2 symmetric tensor. So the degrees of freedom of $I$ must all transform under rotations as rank-2 symmetric tensors, which means that they must lie in the $S \oplus E$ representation of the orthogonal group. This implies that $I$ can only have nonzero $S$ and $E$ components in the Ricci decomposition, and its Weyl component $C$ must vanish. We therefore could have concluded that $C = 0$ directly from equation~\eqref{IKN} without explicitly working out any of the traces of $I$ or the Schouten tensor (although in this case, doing so is not difficult).

Equation~\eqref{IKN} implies a high redundancy in the components of the inertia tensor $I$ that simply reflects the fact that much of the ``work'' that $I$ is doing is simply matching up indices correctly in the tensor contraction. If an algebraic curvature tensor on $\Rn$ has vanishing Weyl component -- or equivalently, if it can be expressed in the form $A \owedge \delta$ -- then its action \eqref{L} simply maps the bivector $\omega$ to twice the antisymmetric part of the matrix product $A \omega$. So in order to calculate the angular momentum $L$ corresponding to an explicit angular velocity bivector $\omega$, the most efficient course of action is often to entirely skip calculating $I$ and to directly use equation~\eqref{Ldirect}.

In three dimensions, the standard 3D inertia tensor~\eqref{I3D} can be expressed in terms of the Schouten tensor~\eqref{A} by
\[
I^{(3D)} = \Tr(A) \delta - A.
\]

The phenomenology of general relativity changes qualitatively between $n \leq 3$ and $n \geq 4$ spacetime dimensions. If $n \leq 3$, then the Weyl tensor field vanishes (or is undefined), so no gravitational degrees of freedom can propagate locally through a vacuum. But if $n \geq 4$, then the possibility of a nontrivial Weyl tensor field enables much richer phenomenology, such as gravitational waves. We have shown that in arbitrary dimensions, the inertia tensor for rigid-body rotation is an algebraic curvature tensor that shares many mathematical similarities with the Riemann curvature tensor. From this parallel alone, we might have guessed (by analogy with general relativity) that the possibility of a nonzero Weyl tensor in the inerta tensor qualitatively changes the phenomenology of rigid-body dynamics in $n \geq 4$ dimensions. But we have shown that this guess is not true, because there is a crucial difference between the Riemann and the inertia tensors: the latter cannot contain a Weyl tensor even in $n \geq 4$ dimensions.\footnote{Of course, there is a much more obvious difference between the Riemann and the inertia tensors: the former is a tensor \emph{field} that varies over a spacetime manifold, while the latter is just a fixed tensor. So the analogy is only rough, and this guess may not have been very plausible in the first place.

Moreover, we are not claiming that there are no qualitative differences between rigid-body dynamics in $n \leq 3$ and $n \geq 4$ dimensions. There are --- most notably, the fact that in $n \geq 4$ dimensions there exist non-simple bivectors, which generate proper rotations that are not confined within a single plane. (In four dimensions, these are sometimes called \emph{double rotations}.) We are only making the narrower claim that -- in contrast with general relativity -- \emph{Weyl tensors} do not cause qualitatively new phenomenology in higher dimensions.}

\begin{acknowledgments}
The author thanks Brayden Ware and Gavin Hartnett for helpful discussions.
\end{acknowledgments}

\bibliography{Inertia}

\begin{thebibliography}{10}%
\makeatletter
\providecommand \@ifxundefined [1]{%
 \@ifx{#1\undefined}
}%
\providecommand \@ifnum [1]{%
 \ifnum #1\expandafter \@firstoftwo
 \else \expandafter \@secondoftwo
 \fi
}%
\providecommand \@ifx [1]{%
 \ifx #1\expandafter \@firstoftwo
 \else \expandafter \@secondoftwo
 \fi
}%
\providecommand \natexlab [1]{#1}%
\providecommand \enquote  [1]{``#1''}%
\providecommand \bibnamefont  [1]{#1}%
\providecommand \bibfnamefont [1]{#1}%
\providecommand \citenamefont [1]{#1}%
\providecommand \href@noop [0]{\@secondoftwo}%
\providecommand \href [0]{\begingroup \@sanitize@url \@href}%
\providecommand \@href[1]{\@@startlink{#1}\@@href}%
\providecommand \@@href[1]{\endgroup#1\@@endlink}%
\providecommand \@sanitize@url [0]{\catcode `\\12\catcode `\$12\catcode
  `\&12\catcode `\#12\catcode `\^12\catcode `\_12\catcode `\%12\relax}%
\providecommand \@@startlink[1]{}%
\providecommand \@@endlink[0]{}%
\providecommand \url  [0]{\begingroup\@sanitize@url \@url }%
\providecommand \@url [1]{\endgroup\@href {#1}{\urlprefix }}%
\providecommand \urlprefix  [0]{URL }%
\providecommand \Eprint [0]{\href }%
\providecommand \doibase [0]{https://doi.org/}%
\providecommand \selectlanguage [0]{\@gobble}%
\providecommand \bibinfo  [0]{\@secondoftwo}%
\providecommand \bibfield  [0]{\@secondoftwo}%
\providecommand \translation [1]{[#1]}%
\providecommand \BibitemOpen [0]{}%
\providecommand \bibitemStop [0]{}%
\providecommand \bibitemNoStop [0]{.\EOS\space}%
\providecommand \EOS [0]{\spacefactor3000\relax}%
\providecommand \BibitemShut  [1]{\csname bibitem#1\endcsname}%
\let\auto@bib@innerbib\@empty
\bibitem [{\citenamefont {Thornton}\ and\ \citenamefont
  {Marion}(2004)}]{Thornton}%
  \BibitemOpen
  \bibfield  {author} {\bibinfo {author} {\bibfnamefont {S.~T.}\ \bibnamefont
  {Thornton}}\ and\ \bibinfo {author} {\bibfnamefont {J.~B.}\ \bibnamefont
  {Marion}},\ }\href@noop {} {\emph {\bibinfo {title} {Classical Dynamics of
  Particles and Systems}}},\ \bibinfo {edition} {5th}\ ed.\ (\bibinfo
  {publisher} {Brooks/Cole},\ \bibinfo {year} {2004})\BibitemShut {NoStop}%
\bibitem [{\citenamefont {Nakahara}(2003)}]{Nakahara}%
  \BibitemOpen
  \bibfield  {author} {\bibinfo {author} {\bibfnamefont {M.}~\bibnamefont
  {Nakahara}},\ }\href@noop {} {\emph {\bibinfo {title} {Geometry, Topology,
  and Physics}}},\ \bibinfo {edition} {2nd}\ ed.,\ Graduate Student Series in
  Physics\ (\bibinfo  {publisher} {Taylor \& Francis Group},\ \bibinfo {year}
  {2003})\BibitemShut {NoStop}%
\bibitem [{\citenamefont {{Math Overflow}}(2011)}]{MO}%
  \BibitemOpen
  \bibfield  {author} {\bibinfo {author} {\bibnamefont {{Math Overflow}}},\
  }\href
  {https://mathoverflow.net/questions/54343/is-there-a-preferable-convention-for-defining-the-wedge-product}
  {\bibinfo {title} {Is there a preferable convention for defining the wedge
  product?}} (\bibinfo {year} {2011})\BibitemShut {NoStop}%
\bibitem [{\citenamefont {Jensen}\ and\ \citenamefont {Poling}(2022)}]{Jensen}%
  \BibitemOpen
  \bibfield  {author} {\bibinfo {author} {\bibfnamefont {S.}~\bibnamefont
  {Jensen}}\ and\ \bibinfo {author} {\bibfnamefont {J.}~\bibnamefont
  {Poling}},\ }\href {https://doi.org/10.48550/ARXIV.2207.03560} {\bibinfo
  {title} {Teaching rotational physics with bivectors}} (\bibinfo {year}
  {2022})\BibitemShut {NoStop}%
\bibitem [{\citenamefont {Berrondo}\ \emph {et~al.}(2012)\citenamefont
  {Berrondo}, \citenamefont {Greenwald},\ and\ \citenamefont
  {Verhaaren}}]{Berrondo}%
  \BibitemOpen
  \bibfield  {author} {\bibinfo {author} {\bibfnamefont {M.}~\bibnamefont
  {Berrondo}}, \bibinfo {author} {\bibfnamefont {J.}~\bibnamefont
  {Greenwald}},\ and\ \bibinfo {author} {\bibfnamefont {C.}~\bibnamefont
  {Verhaaren}},\ }\href {https://doi.org/10.1119/1.4734014} {\bibfield
  {journal} {\bibinfo  {journal} {American Journal of Physics}\ }\textbf
  {\bibinfo {volume} {80}},\ \bibinfo {pages} {905} (\bibinfo {year} {2012})},\
  \Eprint {https://arxiv.org/abs/https://doi.org/10.1119/1.4734014}
  {https://doi.org/10.1119/1.4734014} \BibitemShut {NoStop}%
\bibitem [{\citenamefont {Bini}\ \emph {et~al.}(2003)\citenamefont {Bini},
  \citenamefont {Cherubini}, \citenamefont {Jantzen},\ and\ \citenamefont
  {Ruffini}}]{Bini}%
  \BibitemOpen
  \bibfield  {author} {\bibinfo {author} {\bibfnamefont {D.}~\bibnamefont
  {Bini}}, \bibinfo {author} {\bibfnamefont {C.}~\bibnamefont {Cherubini}},
  \bibinfo {author} {\bibfnamefont {R.~T.}\ \bibnamefont {Jantzen}},\ and\
  \bibinfo {author} {\bibfnamefont {R.}~\bibnamefont {Ruffini}},\ }\href
  {https://doi.org/10.1142/S0218271803003785} {\bibfield  {journal} {\bibinfo
  {journal} {International Journal of Modern Physics D}\ }\textbf {\bibinfo
  {volume} {12}},\ \bibinfo {pages} {1363} (\bibinfo {year} {2003})},\ \Eprint
  {https://arxiv.org/abs/https://doi.org/10.1142/S0218271803003785}
  {https://doi.org/10.1142/S0218271803003785} \BibitemShut {NoStop}%
\bibitem [{\citenamefont {Besse}(1987)}]{Besse}%
  \BibitemOpen
  \bibfield  {author} {\bibinfo {author} {\bibfnamefont {A.~L.}\ \bibnamefont
  {Besse}},\ }\href@noop {} {\emph {\bibinfo {title} {Einstein Manifolds}}},\
  Classics in Mathematics\ (\bibinfo  {publisher} {Springer Berlin},\ \bibinfo
  {year} {1987})\BibitemShut {NoStop}%
\bibitem [{\citenamefont {Petrov}(2000)}]{Petrov}%
  \BibitemOpen
  \bibfield  {author} {\bibinfo {author} {\bibfnamefont {A.~Z.}\ \bibnamefont
  {Petrov}},\ }\href@noop {} {\bibfield  {journal} {\bibinfo  {journal}
  {General Relativity and Gravitation}\ }\textbf {\bibinfo {volume} {32}},\
  \bibinfo {pages} {1665–1685} (\bibinfo {year} {2000})}\BibitemShut
  {NoStop}%
\bibitem [{\citenamefont {Srednicki}(2007)}]{Srednicki}%
  \BibitemOpen
  \bibfield  {author} {\bibinfo {author} {\bibfnamefont {M.}~\bibnamefont
  {Srednicki}},\ }\href@noop {} {\emph {\bibinfo {title} {Quantum Field
  Theory}}}\ (\bibinfo  {publisher} {Cambridge University Press},\ \bibinfo
  {year} {2007})\BibitemShut {NoStop}%
\bibitem [{\citenamefont {Kühnel}\ and\ \citenamefont
  {Rademacher}(2008)}]{Kuhnel}%
  \BibitemOpen
  \bibfield  {author} {\bibinfo {author} {\bibfnamefont {W.}~\bibnamefont
  {Kühnel}}\ and\ \bibinfo {author} {\bibfnamefont {H.-B.}\ \bibnamefont
  {Rademacher}},\ }in\ \href@noop {} {\emph {\bibinfo {booktitle} {Recent
  Developments in Pseudo-Riemannian Geometry}}},\ \bibinfo {editor} {edited by\
  \bibinfo {editor} {\bibfnamefont {D.~V.}\ \bibnamefont {Alekseevskiĭ}}\ and\
  \bibinfo {editor} {\bibfnamefont {H.}~\bibnamefont {Baum}}}\ (\bibinfo
  {publisher} {European Mathematical Society},\ \bibinfo {year} {2008})\ pp.\
  \bibinfo {pages} {261--298}\BibitemShut {NoStop}%
\end{thebibliography}%
\bibliographystyle{apsrev4-2}

\end{document}